\documentclass{elsart}
\usepackage{epsfig}

\begin{document}
\parindent=15pt
\begin{frontmatter}
\title{Mutually compensative pseudo solutions
of primary energy spectra in the knee region}
\author{S.V.~Ter-Antonyan}
  \ead{samvel@yerphi.am}	
\address{Yerevan Physics Institute, 2 Alikhanyan Br. Str., 375036 
Yerevan, Armenia}
%---------------------ABSTRACT----------------------------------------
\begin{abstract}
\indent
The problem of the uniqueness of solutions during the evaluation of
primary energy spectra in the knee region using an extensive air
shower (EAS) data set and the EAS inverse approach is investigated. 
It is shown that the unfolding of primary energy spectra in the 
knee region leads to mutually compensative pseudo solutions.  These
solutions may be the reason for the observed disagreements in the
elementary energy spectra of cosmic rays in the 1-100 PeV energy range
obtained from different experiments.
\end{abstract}
\end{frontmatter}
%-------------------INTRODUCTION and MOTIVATION------------------------
\section{Introduction}
The Extensive air shower (EAS) inverse approach to a problem of the primary
energy spectra reconstruction in the region of $1-100$ PeV energies has been
an essential tool in the past decade 
\cite{Ralph,Samo1,KAS01,Samo2,Samo3,KAS05,Gamma52}. 
Basically, it follows from the high accuracies of recent
experiments \cite{EAS-TOP,CASA,KAS1,Gamma51,Gamma7}
and the availability of the EAS simulation code \cite{CORSIKA},
which was developed in the framework of contemporary interaction models
in order to  
compute the kernel functions of a corresponding integral equation
set \cite{KAS05,Gamma51}. 
At the same time, the energy spectra of primary ($H,He$ and $Fe$)
nuclei obtained from the KASCADE experiment \cite{KAS05} 
using the EAS inverse
approach disagree with the same data from the ongoing GAMMA
experiment \cite{Gamma51,Gamma7}, where parameterization of the EAS
inverse problem is used.\\
\indent
Below, a peculiarity of the EAS inverse problem is investigated, and
one of the possible reasons for the observed disagreements
between the energy spectra in \cite{KAS05}  and \cite{Gamma51} is 
considered in the framework of the SIBYLL \cite{SIB} interaction model.

\indent
The paper is organized as follows: In Section~2 
the EAS inverse approach and the definition of 
the problem of uniqueness is described. 
It is shown, that the abundance of primary nuclear species leads
to pseudo solutions for unfolded primary energy spectra.  
The existence and significance of the
pseudo solutions are shown in Section~4. The pseudo solutions
for primary energy spectra were obtained on the basis of 
simulation of KASCADE \cite{KAS05} shower spectra. 
The EAS  simulation model is presented in Section~3.
In Section 5 the peculiarities of the
pseudo solutions are discussed in comparison with the methodical
errors of the KASCADE data.
%---------------------PROBLEM OF UNIQUENESS---------------------------
\section{Problem of uniqueness}
The EAS inverse problem is
ill-posed by definition and the unfolding of the 
corresponding integral
equations does not ensure the uniqueness of the solutions.
The regularized unfolding on the basis of \textit{a priori} information 
on expected solutions (smoothness, monotony and non-negativity)
in some cases can redefine the inverse problem \cite{Gold}
and provide the appropriate solutions. However, the expected 
singularities (e.g. knees) in the primary energy spectra at 
$10^{15}-10^{16}$ eV may
erroneously be smoothed by regularization algorithms and vice versa,
be imitated by the unavoidable oscillations \cite{Gold} of the solutions.
Furthermore, the EAS inverse problem implies evaluations of at least 
two or more unknown primary energy spectra
from the integral equation
set of Fredholm kind \cite{KAS05,Gamma51,Gamma7}. 
These peculiarities have not been studied in detail
and the problem of the uniqueness of solutions can limit the
number of evaluated spectra.\\
\indent  
Let $f_A(E)$ be the energy spectrum of a primary nucleus $A$ over the
atmosphere, $W_A(\mathbf{x}|E)$ be the probability density
function describing the transformation of $A$ and $E$ parameters of
the primary nucleus to a measurable vector $\mathbf{x}$. 
Then the EAS inverse problem, i.e. the reconstruction 
of the energy spectra of $N_A$ primary nuclei on the basis of 
the detected spectra  $Y(\mathbf{x})$ of EAS parameters, 
is defined by the integral equation 
\begin{equation}
Y(\mathbf{x})=\sum_{A=A_1}^{A_{N_A}}\int f_A(E)W_A(\mathbf{x}|E)\d E\;.
\end{equation}
\indent
Evidently, if $f_{A_1,\dots A_{N_A}}(E)$ are the solutions of eq.~(1),
the functions $f_A(E)+g_A(E)$ should also be the solutions of (1),
provided equation
\begin{equation}
\sum_A\int g_A(E)W_A(\mathbf{x}|E)\d E 
=0(\pm\Delta Y)
\end{equation}
is satisfied for the given measurement errors $\Delta Y(\mathbf{x})$
and for at least one of the combinations of the primary nuclei 
\begin{equation}
n_C=\sum_{j=1}^{N_A}{N_A \choose j}\;. 
\end{equation}
The number of combinations (3) stems from a
possibility of the existence of a set of functions
$g_A(E)\equiv g_{1,A}(E),\dots,g_{i,A}(E)$ for
each of the primary nuclei ($A$), which can independently
satisfy eq.~(2).\\
\indent
For example, suppose that $N_A=3$. 
Let us denote 
$\int g_{i,A_k}(E)W_{A_k}(E)\d E$ by $I_{i,A_k}$ 
and, for simplicity, set the right-hand side of eq.~(2) to 0.
Then, following expression (3), we find $n_C=7$ independent combinations 
of eq.~(2):
$I_{1,A_k}=0$ for $k=1,2$ and $3$, 
$I_{2,A_1}+I_{2,A_2}=0$, $I_{3,A_1}+I_{2,A_3}=0$,
$I_{3,A_2}+I_{3,A_3}=0$ and $I_{4,A_1}+
I_{4,A_2}+I_{4,A_3}=0$ with different 
$g_{i,A_k}(E)$ functions. 
The measurement errors $\pm\Delta Y$ on the right-hand side 
of these equations can both increase and decrease the
domains of  $g_{i,A_k}(E)$ functions.\\
\indent
One may call the set of functions $g_A(E)$ the
pseudo functions with the corresponding pseudo solutions (spectra)
$f_A(E)+g_A(E)$. The oscillating $g_A(E)\equiv g_{1,A}(E)$
functions at $j=1$ are 
responsible for the first $N_A$ equations
$\int g_{1,A}(E)W_A(\mathbf{x}|E)\d E=0(\pm\Delta Y)$,
$A\equiv A_1,\dots A_{N_A}$,
due to the positive-definite
probability density function $W_A(E)$.
The pseudo solutions $f_A(E)+g_{1,A}(E)$ can be avoided by 
using iterative unfolding algorithms \cite{KAS05,Gold}.\\
\indent
Additional sources of the pseudo solutions 
originate from the mutually compensative effects at $j\geq2$:
\begin{equation}  
-\sum_k\int g_{A_k}(E)W_{A_k}(\mathbf{x}|E)\d E\simeq
\sum_{m\neq k}\int g_{A_m}(E)W_{A_m}(\mathbf{x}|E)\d E
\end{equation}
inherent to eq.~(2) for arbitrary groups of $k$ 
and $m\neq k$ primary nuclei. Since there are no 
limitations on the types of the pseudo functions
(except for $f_A(E)+g_A(E)>0$) that would follow from expression (4),
and the number of
possible combinations (3) rapidly increases with the number of
evaluated primary spectra ($N_A$), 
the problem of the uniqueness of solutions may be 
insoluble for $N_A>3$. 
Moreover, the pseudo functions
have to restrict the efficiency of
unfolding energy spectra for $N_A\simeq2-3$, because
the unification of $Z=1,\dots,28$ primary nuclei
spectra into $2-3$ nuclear species (e.g. light and heavy)
inevitably increases the
uncertainties of the kernel functions $W_A(E)$
and thereby also increases the domains 
of the pseudo functions.\\
\indent
Notice, that the pseudo solutions will always appear in the
iterative unfolding algorithms if the initial iterative values
are varied within large intervals. 
At the same time, it is practically impossible
to derive the pseudo functions from the unfolding of 
equations~(1,2) due to a strong ill-posedness of the inverse problem.
However, for a given set of the measurement errors 
$\Delta Y(\mathbf{x})$ and the known kernel functions 
$W_A(\mathbf{x}|E)$ for $A\equiv A_1,\dots A_{N_A}$
primary nuclei, eq.~(2) can be regularized
by parametrization of the pseudo functions 
$g_A(\alpha,\beta,\dots|E)$.
The unknown parameters $(\alpha,\beta,\dots)$ can be
derived from a numerical solution of parametric eq.~(2), and
thereby one may also evaluate the parametrized
pseudo functions $g_A(E)$.\\  
\indent
Below (Section~3), an EAS simulation model for computing the 
kernel function $W_A(E)$ and replicating the KASCADE \cite{KAS05}
EAS spectral errors $\Delta Y(\mathbf{x})$ is considered.  
%---------------------EAS SIMULATION MODEL---------------------
\section{EAS simulation model} 
The primary energy spectra obtained in the KASCADE experiment 
were derived on the basis 
of the detected 2-dimensional EAS size spectra 
$Y(\mathbf{x})\equiv Y(N_e,N_{\mu})$ and an iterative
unfolding algorithm \cite{Gold}
for $N_A=5$ primary nuclei \cite{KAS05}. 
Evidently, whether these solutions are unique or not depends on
the significance of the arbitrary pseudo functions 
$|g_A(E)|$ from eq.~(2).\\
\indent
We suppose that the convolution of the shower spectra 
$W_A(N_e,N_{\mu}|E)$ at the observation level
and corresponding measurement errors $\sigma(N_e)$, $\sigma(N_{\mu})$
\cite{Ralph}
are described by 2-dimensional log-normal distributions with
parameters
$\xi_e=\overline{\ln{N_e}}(A,E)$, 
$\xi_{\mu}=\overline{\ln{N_{\mu}}}(A,E)$,
$\sigma_e(A,E)$, $\sigma_{\mu}(A,E)$ 
and correlation coefficients 
$\rho_{e,{\mu}}(A,E)$ between the shower size 
($\ln N_e$) and  the muon truncated size ($\ln N_{\mu}$). 
We tested this hypothesis
by the $\chi^2$ goodness-of-fit test using 
the CORSIKA(NKG) EAS simulation
code \cite{CORSIKA} for the SIBYLL2.1 \cite{SIB} interaction model, 
4 kinds of primary nuclei ($A\equiv p,He,O,Fe$),
5 energies ($E\equiv 1,3.16,10,31.6,100$ PeV) 
and simulation samples for each of $E$ and $A$:
5000, 3000, 2000, 1500, 1000 respectively in $0-18^0$
zenith angular interval.
The values of corresponding 
$\chi^2(A_i,E_j)/n_{d.f.}$, ($i=1,\dots4$, $j=1,\dots5$) were 
distributed randomly in the interval $0.5-1.4$ 
for the measurement ranges of the KASCADE experiment 
($N_{e,\min}=6.3\cdot10^4$ 
and $N_{\mu,\min}=4\cdot10^3$)
and the bin sizes $\Delta\ln N_e,\Delta\ln N_{\mu}=0.075$.\\
\indent
Notice, that the combined 2-dimensional log-normal distributions
with parameters
$\sigma_{e,1}(A,E)$ at $\ln N_e<\xi_e$,
$\sigma_{e,2}(A,E)$ at $\ln N_e>\xi_e$,
$\sigma_{\mu,1}(A,E)$ at $\ln N_{\mu}<\xi_{\mu}$ and 
$\sigma_{\mu,2}(A,E)$ at $\ln N_{\mu}>\xi_{\mu}$,
more precisely ($\chi^2/n_{d.f.}\le1.2$) describe the shower spectra 
$W_A(N_e,N_{\mu}|E)$ in the tail regions.\\
\indent
We performed an additional test of the log-normal
fit of the $W_A$ spectra using
multiple correlation analysis for the shower parameters 
simulated by the 
log-normal $W_A(N_e,N_{\mu}|E)$ probability density functions
and shower parameters obtained from the CORSIKA EAS simulations 
at power-law primary energy spectra ($\gamma=-1.5$) and equivalent
abundances of primary nuclei. 
The corresponding correlation coefficients were equal to
$\rho(\ln E|\ln N_e,\ln N_{\mu})=0.97$, 
$\rho(\ln A|\ln N_e,\ln N_{\mu})=0.71$, 
$\rho(\ln A,\ln N_e)=-0.14\pm0.01$, 
$\rho(\ln A,\ln N_{\mu})=0.18\pm0.01$, 
and were in close agreement for both methods 
of $N_e$ and $N_{\mu}$ generations.

\indent
We replicated the KASCADE 2-dimensional EAS size spectrum 
$Y(N_e,N_{\mu})$ (and corresponding $\Delta Y$ )
by picking out $N_e$ and $N_\mu$
randomly from the 2-dimensional shower spectra 
$W_A(N_e,N_{\mu}|E)$ after randomly picking $A$ and $E$ 
parameters of a primary particle from the power-law energy spectra
\begin{equation}
f_A(E)\propto E^{-2.7}
\Big(1+\Big(\frac{E}{E_k}\Big)^\epsilon\Big)^{-0.5/\epsilon}
\end{equation}
with a rigidity-dependent knee $E_k=Z\cdot2000TV$, the sharpness 
parameter $\epsilon=3$ and normalization of the all-particle
spectrum $\int\sum_{A}f_A(E)\d E=1$.
The relative abundance of nuclei was arbitrarily chosen to be
$0.3,0.45,0.15$ and $0.1$ for primary $H,He,O$ and $Fe$ nuclei
respectively,
which approximately conforms with the expected abundance 
from balloon and satellite data \cite{Wiebel}.\\ 
The mediate values of the parameters of the 
probability density function
$W_A(N_e,N_m|E)$ 
were estimated by the corresponding log-parabolic splines.\\
\indent
The total number of 
simulated EAS events was set to $7\cdot10^5$ 
in order to replicate the corresponding statistical errors 
$\Delta Y(N_e,N_{\mu})$ of the KASCADE data.
%---------------------PSEUDO SOLUTIONS---------------------
\section{Pseudo solutions} 
On the basis of the obtained estimations of $\Delta Y(N_e,N_{\mu})$
(Section ~3) 
for the KASCADE experiment, we examined the uniqueness of unfolding (1)
by $\chi^2$-the minimization:
\begin{equation}
\chi^2=\sum_{i=1}^{I}\sum_{j=1}^{J}
\bigg(\frac{G(N_{e,i},N_{\mu,j})}
{\Delta Y(N_{e,i},N_{\mu,j})}\bigg)^2\;,
\end{equation}
where $G(N_{e,i},N_{\mu,j})$ represents the left-hand side of
eq.~(2) for 2 kinds of the empirical pseudo functions
\begin{equation}
g_A(E)=\alpha_A\Big(\frac{E}{E_m}\Big)^{-\gamma_A}\;,
\end{equation}
\begin{equation}
g_A(E)=\alpha_A((\ln E-\beta_A)^3+\eta_A)\Big(\frac{E}{E_m}
\Big)^{-3}\;,
\end{equation}
while $g_A(E)+f_A(E)>0$, otherwise $g_A(E)=-f_A(E)$. 
The unknown 
$\alpha_A,\beta_A,\gamma_A$ and $\eta_A$
parameters in expressions (7,8)
were derived from $\chi^2$ minimization (6).
The numbers of bins were $I=60$ and $J=45$
with the bin size 
$\Delta \ln{N_e},\Delta\ln{N_{\mu}}\simeq0.1$.\\  
\indent
In fact, the minimization of $\chi^2$ (6) 
for different representations
(7,8) of the pseudo functions $g_A(E)$
provides a solution of the corresponding parametric
eq.~(2) with a zero right-hand side.
To avoid the trivial solutions $g_A(E)\equiv0$
and reveal the domains of the pseudo functions,
the values of some of the  parameters
were arbitrarily 
fixed during the minimization of $\chi^2$ (6).
The magnitudes of the fixed parameters were empirically
determined via optimization of conditions 
$\chi^2_{\min}/n_{d.f.}\simeq1$ and $|g_A(E)|\sim f_A(E)$ for the
pseudo spectra with the fixed parameters.\\
\indent
The true primary energy spectra 
$f_A(E)$ for $A\equiv H,He,O,Fe$ nuclei (5) and the all-particle
energy spectrum $\sum f_A(E)$ (lines) along with the 
corresponding distorted (pseudo) spectra $f_A(E)+g_A(E)$
(symbols) are presented in Fig.~1 respectively.
\begin{figure}[h]  %--------------------------fig.1-----------------
\vspace{5mm}
\begin{center}
\includegraphics[width=7cm,height=7cm]{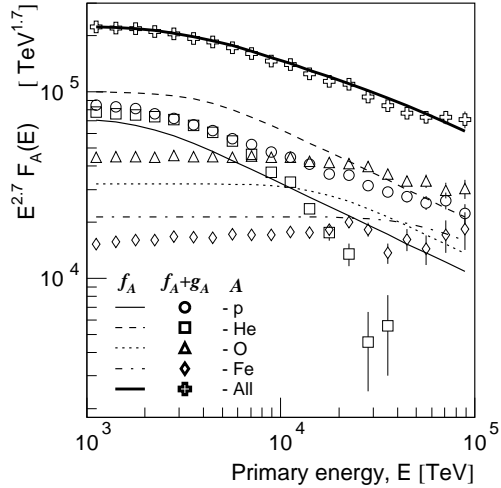}
\end{center}
\caption{Primary energy spectra $f_A(E)$ and the 
all-particle spectrum
$\sum f_A(E)$ for $A\equiv H,He,O,Fe$ nuclei (lines) 
and the corresponding pseudo solutions $f_A(E)+g_A(E)$ 
for the pseudo function (7) (symbols).}
\vspace{10mm}
\end{figure}
The parameters of the pseudo functions (7) derived
for  $\chi^2_{\min}/n_{d.f.}=1.08$ ($n_{d.f.}=717$) 
are presented in Table~1.\\ 
\begin{table}  %******************************Table 1****************
\vspace{5mm} 
\caption{Parameters $\alpha_A$ (TeV$^{-1}$) and $\gamma_A$ 
of the pseudo function (7) for different primary nuclei $A$
and $E_m=1000$ TeV.}
\begin{center}
\begin{tabular}{lcc}
\hline
$A $& $ \alpha_A\cdot10^4  $&$  \gamma_A    $\\
\hline
$p $& $1.10\pm0.06    $&$2.71\pm0.04$\\
$He$& $-1.80$ (fixed)  &$2.60$ (fixed)\\
$O $& $0.97\pm0.05    $&$2.65\pm0.04$\\
$Fe$& $-0.50$ (fixed)  &$2.90$ (fixed)\\
\hline
\end{tabular}
\end{center}
\vspace{10mm}
\end{table}
\indent
The effect of the pseudo functions (8) on the resulting primary
energy spectra is shown in Fig.~2. 
\begin{figure}[h]  %--------------------------fig.2-----------------
\vspace{5mm} 
\begin{center}
\includegraphics[width=7cm,height=7cm]{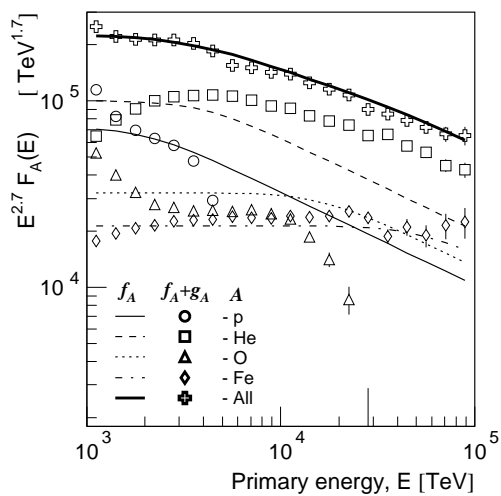}
\end{center}
\caption{The same as Fig.~1 for the pseudo function (8).}
\vspace{10mm}
\end{figure}
Evaluations of the corresponding 
parameters are presented in Table~2 for
$\chi^2_{\min}/n_{d.f.}=1.1$.
\begin{table}  %******************************Table 2****************
\vspace{5mm} 
\caption{Parameters $\alpha_A$ (TeV$^{-1}$), $\gamma_A$ and $\eta$ 
of the pseudo function (8) for different primary nuclei $A$
and $E_m=1000$ TeV.}
\begin{center}
\begin{tabular}{lccc}
\hline
$A $& $ \alpha_A\cdot10^4  $&$  \beta_A    $&$\eta_A$\\
\hline
$p $& $-9.00$  (fixed) &$7.76\pm0.01$   &$0$ (fixed)\\
$He$& $0.044\pm0.02   $&$13.2\pm1.08$   &$169\pm98$ \\
$O $& $-0.80$  (fixed) &$8.47\pm0.05$   &$0.94\pm0.16$\\
$Fe$& $ 0.010\pm0.002 $&$11.4\pm0.14$   &$50$ (fixed)\\
\hline
\end{tabular}
\end{center}
\vspace{10mm}
\end{table}
The variations of the cubic power indices in expression (8)
in the range of $2-5$ lead
to different types
of pseudo solutions as well.\\
\indent
It is clear from Figs.~1,2, that the contribution of the pseudo 
functions $g_A(E)$
can be comparable and even significantly 
larger than the values of the true spectra $f_A(E)$.
Moreover, the pseudo solutions lose both the slopes  
and the intensities of the spectra. At the same time, the
all-particle spectra slightly depend on the
contribution of the pseudo functions.\\
\indent
The same results (Tables~1,2)
were obtained using both the combined 2-dimensional
log-normal representation of the shower spectra 
$W_A(N_e,N_{\mu}|E)$
(Section 3) and the 3-dimensional 
($\ln E,\ln N_e,\ln N_{\mu}$) parabolic 
interpolations of corresponding probability density 
functions obtained by the CORSIKA code.\\
\indent
Evidently, the range of relatively large
measurement errors $\Delta Y(\mathbf{x})$ expands the domain 
of the pseudo functions.  Contributions 
of the mutually compensative effects (eqs.~2,4) of the pseudo functions
to the domain of the pseudo solutions
were tested using a 10 times larger 
EAS simulation sample ($n=7\cdot10^6$) and the pseudo functions
with evident singularity:
\begin{equation}
g_A(E)=\alpha_A\varepsilon_A^{-1}
\Big(\frac{E}{\varepsilon_A}\Big)^{\delta} \;,
\end{equation}
where $\delta=-1$ at $E\leq\varepsilon_A$
and $\delta=-7$ at $E>\varepsilon_A$.
The singularity of the pseudo function (9) for $A\equiv H$ was fixed 
at $\varepsilon_H=3000$ TeV 
and the scale factor $\alpha_H=-0.03$.
The remaining parameters for 
primary nuclei $A\equiv He,O,Fe$ 
were estimated by $\chi^2$-minimization (6)
and presented in Table~3 for $\chi^2_{\min}/n_{d.f.}=2.01$
and $n_{d.f.}=857$. The accuracies of integrations (2) were
about $0.1\%$.
The corresponding pseudo solutions are shown in Fig.~3.\\
\begin{table}  %************************ Table 3 ****************
\vspace{5mm} 
\caption{Parameters $\alpha_A$ (TeV$^{-1}$) and $\varepsilon_A$ (TeV) 
of the pseudo function (9) for different primary nuclei $A$
and $\varepsilon_H=3000$ TeV.}
\begin{center}
\begin{tabular}{lcc}
\hline
$A $& $ \alpha_A\cdot100  $&$  \varepsilon_A/\varepsilon_H   $\\
\hline
$p $& $-3.0$ (fixed)  &$1$ (fixed)\\
$He$& $3.05\pm0.07$   &$1.03\pm0.01$\\
$O $& $-0.84\pm0.06$  &$1.08\pm0.03$\\
$Fe$& $ 0.15\pm0.02$  &$1.29\pm0.10$\\
\hline
\end{tabular}
\end{center}
\vspace{10mm}
\end{table}
\begin{figure}[h]  %--------------------------fig.3-----------------
\vspace{5mm} 
\begin{center}
\includegraphics[width=7cm,height=7cm]{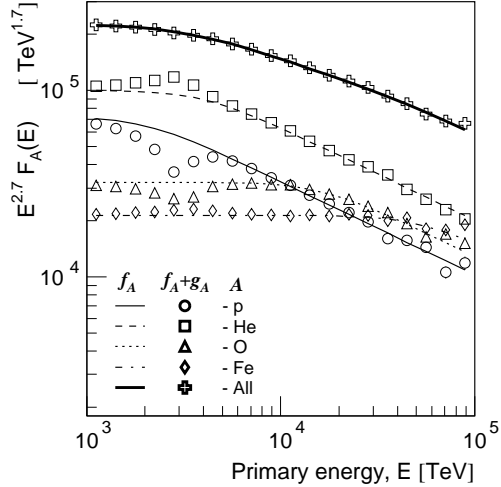}
\end{center}
\caption{The same as Fig.~1 for the pseudo function (9) and $n=7\cdot10^6$
simulated showers.}
\vspace{10mm}
\end{figure}
\indent
Since the measurement errors are negligibly small,
the significance of the mutually compensative 
effects is well seen. The singularity of the proton spectrum 
was approximately compensated by the $He$ and $O$ spectra.  
This is due to both the large number ($n_C=15$) 
of possible mutually compensative combinations (3) and
the peculiarities of EAS development 
in the atmosphere (kernel functions $W_A(E)$, Section~3), 
which
are expressed by the approximately log-linear
dependences of the statistical parameters
$<\ln{N_e}>$, $<\ln{N_{\mu}}>$, $\sigma_e$ and $\sigma_{\mu}$
of shower spectra $W_A(E)$ on energy ($\ln{E}$)
and nucleon number ($\ln{A}$) of primary nuclei 
\cite{Samo7,Hoerandel2}. The value of 
$\chi^2_{\min}/n_{d.f.}$ for a $10$ times
smaller EAS sample ($n=7\cdot10^5$) was equal to $0.25$.
%--------------------- DISCUSSION ---------------------
\section{Discussion}
The results from Figs.~1--3
show that the pseudo functions with mutually
compensative effects exist and belong  
practically to all families - linear~(7), 
non-liner~(8) and even singular~(9) 
in a logarithmic scale.\\
\indent
The all-particle energy
spectra in Figs.~1--3 
are practically indifferent to the
pseudo solutions of elemental spectra.
This fact directly follows from eq.~(2)
for pseudo solutions
and is well confirmed 
by the identity of the GAMMA \cite{Gamma51,Gamma7} 
and KASCADE \cite{KAS05} all-particle energy spectra in spite of 
disagreements of the elemental 
($p,He,Fe$) primary energy spectra (see \cite{Gamma51,Gamma7}).\\
\indent
The $\chi^2$ minimization (6)
uses mainly the nearest pseudo energy spectra
with free parameters for compensation of 
the pseudo spectra with fixed 
parameters.
\begin{figure}[h]  %--------------------------fig.4-----------------
\vspace{5mm} 
\begin{center}
\includegraphics[width=7cm,height=7cm]{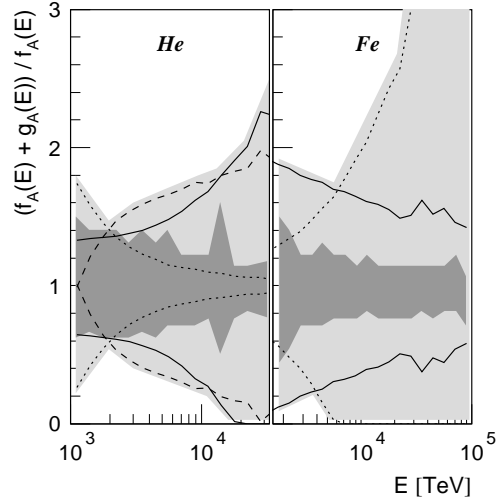}
\end{center}
\caption{Domains of the pseudo solutions 
for $He$ and $Fe$ primary nuclei (light shaded
areas) and corresponding  "methodical
errors" of the KASCADE unfolding spectra 
(dark shaded areas) taken from \cite{KAS05}. 
The solid and dotted lines resulted from
pseudo functions (7) and the dashed lines stemmed
from (8).}
\vspace{10mm}
\end{figure}
The significance of the pseudo functions
$|g_A(E)|$ in most cases exceeds 
the significance of the evaluated primary energy 
spectra $f_A(E)$ and unfolding of (1) 
can not be effective for $N_A=4$.\\
\indent
The unfolding of the primary energy spectra for $N_A=5$
will increase the number of possible combinations (3)
of the pseudo solutions and the corresponding 
pseudo functions by a factor of two.
Taking into account the large values
of applied $\chi_{\min}^2/n_{d.f.}\simeq2-3$ \cite{KAS05}
one may conclude that the contributions
of the pseudo functions in the unfolded energy
spectra of \cite{KAS05} have to be dominant.\\
\indent
The "methodical errors" obtained in \cite{KAS05} 
for $N_A=5$ define  
the uncertainties of the solutions intrinsic 
only to the given unfolding algorithms.  
The existence and significance of 
the mutually compensative pseudo solutions 
follow from eqs.~(1,2)
and from the peculiarities of the
shower spectra $W_A(\mathbf{x}|E)$ 
regardless of the unfolding algorithms.\\
\indent
Comparison of the methodical errors
$(f_A(E)+\Delta f_A(E))/f_A(E)$ for $A\equiv He$ 
and $A\equiv Fe$ from \cite{KAS05}
with corresponding errors $(f_A(E)+g_A(E))/f_A(E)$ 
due to the pseudo solutions from expressions (7,8) are shown
in Fig.~4. 
The magnitudes of the fixed parameters were 
empirically determined by maximizing $|g_{He}(E)|$
(left panel) and $|g_{Fe}(E)|$ (right panel)
for a given goodness-of-fit test
$\chi^2_{\min}/n_{d.f}\simeq2.5$ from \cite{KAS05}.\\
\indent
It is seen that the methodical
errors (dark shaded areas) from \cite{KAS05} significantly 
underestimate the contribution of the pseudo solutions
(light shaded areas) from expressions(7,8). Moreover, the 
methodical errors from \cite{KAS05} slightly depend on the 
primary energy (or statistical errors), whereas
the domains of the pseudo solutions strongly correlate
with the statistical errors according to definition (2).
%--------------------------CONCLUSION---------------------------------
\section{Conclusion}
The results show that the reconstruction of primary energy
spectra using unfolding algorithms \cite{KAS05,Gold} can not be
effective and the disagreement between the KASCADE \cite{KAS05}
and GAMMA \cite{Gamma51,Gamma7} data is insignificant in 
comparison with the large domains of the
mutually compensative pseudo solutions (Fig~4)
of the unfolded spectra \cite{KAS05}.\\
\indent
Even though the oscillating pseudo solutions
$g_{1,A}(E)$ (Section 2) are possible to avoid using regularization
algorithms \cite{Gold}, the mutually compensative effect (4) of the 
arbitrary pseudo functions $g_{A}(E)$ 
intrinsic to the expression (2) is practically impossible 
to avoid at $N_A>1$.\\
\indent
The uncertainties of solutions 
due to the mutually compensative 
pseudo functions can be obtained by varying the 
initial values of iterations within a wide range
in the frameworks of a given unfolding algorithm.\\
\indent
To decrease the contributions 
of the mutually compensative pseudo solutions
one may apply a parameterization of
the integral equations (1) 
\cite{Ralph,Samo1,Samo2,Gamma51,Gamma7}
using \textit{a priori} (expected from theories 
\cite{Hillas,Hoerandel1,Stanev})
known primary energy spectra with a set of free
spectral parameters. 
This transforms the EAS inverse problem into
a set of equations with unknown spectral parameters,
and thereby the EAS
inverse problem is transmuted into a test of 
the given primary energy
spectra using detected EAS data \cite{Samo2}.
The reliability of the solutions can be 
determined by their stability depending on
the number of spectral parameters, the agreement
between the expected and detected EAS data sets, 
and the conformity of the spectral
parameters with theoretic predictions.\\
\indent
The all-particle energy spectra (Fig.~1--3) 
are practically indifferent toward the  pseudo solutions for 
elemental spectra.\\
The obtained results depend slightly on the spectral representations
of the shower spectra $W_A(E)$ and the primary energy spectra $f_A(E)$.
%------------------ Acknowledgments -------------------------------
\section*{Acknowledgments}
I thank my colleagues from the GAMMA experiment for stimulating
this work and the anonymous referee for suggestions which
considerably improved the paper.
%--------------------------REFERENCE------------------------

\end{document}